\documentstyle[epsf,12pt]{article}
\include{feynman}
\newcommand{\nll}{\nonumber \\}

\newcommand{\bq}{\begin{equation}}
\newcommand{\eq}{\end{equation}}
\newcommand{\ba}{\begin{eqnarray}}
\newcommand{\ea}{\end{eqnarray}}
\newcommand{\nobody}{\rule{0ex}{1ex}}

\begin{document}
\begin{flushright}
MPI-PhT/94-41\\ LMU-10/94\\ July 1994\vspace*{3cm}
\end{flushright}
\begin{center}
{\LARGE\bf Production of heavy bound states\\ at LEP and beyond\footnote
{Talk presented by R. R\"uckl at the Zeuthen Workshop on Elementary Particle
Theory, '{\it Physics at LEP200 and Beyond}', Teupitz, Germany, 10-15 April,
1994}$^,$\footnote{Supported by the German Federal Ministry for Research
and Technology under contract No.~05~6MU93P.}}
\vspace{2cm}\\
Arnd~Leike$^{\rm a}$
        and
Reinhold~R\"uckl$^{\rm a,b}$\vspace{0.5cm}\\
$\nobody^{\rm a}${\small\it
Sektion Physik der Universit\"at M\"unchen,\\
 Theresienstr. 37, D--80333 M\"unchen, FRG}\\
$\nobody^{\rm b}${\small\it
Max-Planck-Institut f\"ur Physik, Werner-Heisenberg-Institut,\\
F\"ohringer Ring~6, D-80805 M\"unchen, FRG}
\vspace*{4cm}\\
{\bf Abstract}
\end{center}
{\small
We describe some characteristic properties of $B_c$ mesons and
discuss the production and the prospects of detection in
$e^+ e^-$, $\gamma \gamma$, and $\bar pp/pp$ collisions. The
production mechanisms considered here also play an important role in
charmonium and bottomonium production.
}
\vfill\newpage
%
\section{INTRODUCTION}
With the production of $10^7$ $Z$ bosons at LEP rare Z decays with branching
ratios as low as $10^{-6}$ come into experimental reach. Among these are
decays into heavy quark bound states,
\bq
Z\rightarrow n^{2S+1} L_J + \mbox{\ anything},
\eq
where the usual spectroscopic notation is used.
Particularly interesting is the possibility to produce bottom--charm bound
states, so--called
$B_c$ mesons. These states have not yet been discovered. They are the only
quarkonium states with open heavy flavour quantum numbers, since the top
quark is expected to be
too heavy to form $T_c$ and $T_b$ resonances before it decays via
$t \rightarrow bW^+$. Moreover, $B_c$ mesons constitute an excellent laboratory
for studying QCD bound state dynamics, weak decay mechanisms, quark and
gluon fragmentation, and hadronic matrix elements. All this provides strong
motivation to investigate the prospects for observing $B_c$ mesons in
present and future experiments.

In the following we briefly describe some
characteristic properties of $B_c$ mesons,
discuss the dominant production mechanisms in $e^+ e^-,\ \gamma\gamma$, and
$\bar pp/pp$ collisions, and indicate the prospects for detection
of these new bound states.
\section{SPECTROSCOPY}
The energy levels and other spectroscopic properties of the $[\bar{b}c]$
system can be calculated in nonrelativistic potential models or with the
help of QCD sum rules. Since the $B_c$ resonances lie in the mass range
between charmonium and bottomonium, the predictions in a given approach
are constrained by the latter systems. Thus,
measurements of $B_c$ properties provide interesting tests of potential
models and sum rule methods.

\begin{center}\begin{tabular}{|c|r|r|r|}
\hline
MeV                & Martin & BT   & moment SR\\
\hline
$m_{B_c}$            & 6247 & 6272 & 6240-6320\\
$m_{B^*_c}-m_{B_c}$  &   72 &   70 &   20-80\\
$f_{B_c}$            &  511 &  442 &  410-440\\
$f_{B^*_c}-f_{B_c}$  &    0 &    0 &    20-40\\
\hline
\end{tabular}
\end{center}
Table 1.
Masses and decay constants \cite{RR} of the $^1S_0$ and $^3S_1$ ground
states from potential models (Martin \cite{M}, BT \cite{BT}) and
moment sum rules \cite{SVZ}.
The normalization is such that $f_\pi=132\,MeV$ and
$f_{B_c^*}=f_{B_c}$ for nonrelativistic $S-$waves.
\vspace{3mm}

In table 1 we exemplify typical
predictions for the pseudoscalar $(1^{1} S_0)$ and vector $(1^{3} S_1)$ ground
states. While the mass predictions are very similar in different models
and approaches, the decay constants are more distinctive (for reviews
see e.g. \cite{K}).
Since the bound
state wave functions are known in nonrelativistic approximation, one can
also calculate more complicated quantities such as transition form factors
and fragmentation functions, for example, for $B_c \rightarrow J/\psi $
and $\bar b \rightarrow B_c \bar c $. We will return to this possibility in
section 4.
\section{WEAK DECAYS}
%
The weak decays of the lightest $B_c$ meson proceed via three distinct
mechanisms: decay of the bottom constituent, decay of the charm constituent,
and $\bar b c$ annihilation. In fact, the free quark lifetimes,
$\tau_{b} \simeq 1.5\,ps$ and $\tau_c \simeq 0.7\,ps$, are quite
comparable, and also the lifetime $\tau_{\bar bc} \simeq\,5.2\,ps$ due to
the annihilation process is not very much longer. The interplay
of these mechanisms should make $B_c$ decays rather unique and give rise to
a rich and interesting decay pattern.

The decay widths corresponding to the
above mechanisms have been estimated in quark models \cite{CHLUS,KBARI} and
using QCD sum rules \cite{KBARI,C}.
For the relative rates one finds
\bq
\Gamma_b : \Gamma_c : \Gamma_{\bar{b}c} \simeq
\left\{\begin{array}{ll} 37 : 45 : 18 & {\rm potentials}\\
                         48 : 39 : 13 & {\rm sum\ rules}.\end{array}\right.
\eq
Deviations from the ratios $29 : 62 : 8$ of the above inverse lifetimes
$\tau ^{-1}_b : \tau ^{-1}_c : \tau ^{-1}_{\bar bc}$ are expected due to
differences in the phase space available for the three decay channels. The
total $B_c$ lifetime is estimated to lie in the range
\bq
\tau_{B_c} = 0.4 \div 0.9\,ps .
\eq

Decay modes with a $J/\psi$ resonance in the final state provide a
particularly favourable signature for detection.
Unfortunately, the branching ratios are only known with considerable
uncertainties. For the most important channels one expects
\bq
\label{branbc}
Br(B_c\rightarrow J/\psi X) =
\left\{\begin{array}{rl}   2\div 5\%   & X=\bar{l}\nu_l\\
                         0.2\div 0.4\% & X=\pi\\
                         0.6\div 1.1\% & X=\rho.     \end{array}\right.
\eq
The inclusive branching ratio for $B_c \rightarrow J/\psi X $ is about 20\%.
It should also be noted that the excited states below the open flavour
threshold, $m_B + m_D \simeq 7.142\,GeV$, will all decay into the $B_c$ ground
state via emission of photons, pions, etc., and hence contribute to the
inclusive $B_c$ rate.
\section{PRODUCTION IN $e^+e^-$ ANNIHILATION\hfill AND
 HEAVY QUARK FRAGMENTATION}
%
In $e^+ e^-$ annihilation $B_c $ mesons emerge from the simultaneous
production of a $b \bar b $ and $c \bar c $ pair of quarks by strong
binding of the $\bar bc$ pair in a colour--singlet state.
(If no confusion can arise, both $B_c$ and $B_c^*$ are simply denoted by
$B_c$.) The lowest--order
Feynman diagrams of this process are shown in Fig.~1. Because of the
nonrelativistic nature of $[\bar bc]$ bound states, the relative momentum $p$
of the $\bar b$ and $c$ quarks is expected to be small in comparison to the
quark masses. It is then reasonable to expand the amplitudes in $p$, and to
keep only the lowest nonvanishing term \cite{GKPR}. In this approximation,
the amplitudes for the production of $S$--waves reduce to hard scattering
amplitudes multiplied by the $S$--wave function $\Psi(0)$ at the origin.
 Similarly, for
$P$--waves one obtains a product involving the derivative of the $P$--wave
function. These wave functions can be estimated from potential models. For
$S$--waves one can also substitute $\Psi(0)$ by the corresponding decay
constants, $|\Psi (0)| = \sqrt{m_{B_c}/12} f_{B_c}$,
and use for the latter the QCD sum rules estimates indicated in Table 1.

Straightforward
calculation of the Feynman diagrams of Fig.~1 \cite{EE} then
yields the total $B_c$ and $B_c^*$ cross sections exhibited in Fig.~2. As
can be seen, below the $Z$ peak the cross sections amount to a few
femtobarns only. At the resonance they reach about $1\div 2\,pb$
corresponding to
$300 \div 500$ events for $10^7$ $Z$ bosons produced. Above the resonance
the cross sections again drop rapidly below the level of 1 fb at LEP200.
The production process considered here dominates other possible mechanisms
such as soft fragmentation, that is nonperturbative creation of a heavy
quark pair from the vacuum, and $W$--radiation from light  fermions
followed by $W^* \rightarrow \bar bc \rightarrow B_c $.
The calculation can be drastically simplified for the following two
reasons: Firstly, the minimum virtuality $k^2$ of the gluons in Fig.~1
is determined by the mass of the secondary quark pair, i.e.
$k^2 \ge 4 m_c^2$ in 1a and b, and $k^2 \ge 4 m_b^2$ in 1c and d.
Therefore, the gauge invariant subset of diagrams, 1a and b, is expected
to yield the dominant contribution. Indeed, at $\sqrt{s} \ge 80 GeV$ the
 diagrams 1c and d contribute less than 5\% as can be seen from
Fig.~3.
Secondly, in a planar gauge such as $Q^{\mu }G_{\mu }= 0$, where
$Q^{\mu } = (p_{e^+} + p_{e^-})^{\mu }$ and $G^{\mu }$ denotes the gluon
field, the contribution from diagram 1b is suppressed relative to that
 from diagram 1a. Again, this is demonstrated in
Fig.~3. Consequently, for $s \gg m^2_{B_c}$
the complete process can be described to a very good approximation as
production and fragmentation of $b$ quarks, $e^+ e^- \rightarrow b \bar b$,\
$\bar b \rightarrow B_c \bar c$.
This feature has been exploited
in \cite{bcfrag,F1,falk}.
\ \vspace{1cm}\\
\begin{minipage}[t]{7.8cm}{
\begin{center}
\hspace{-1.7cm}
\mbox{
\epsfysize=7.0cm
\epsffile[0 0 500 500]{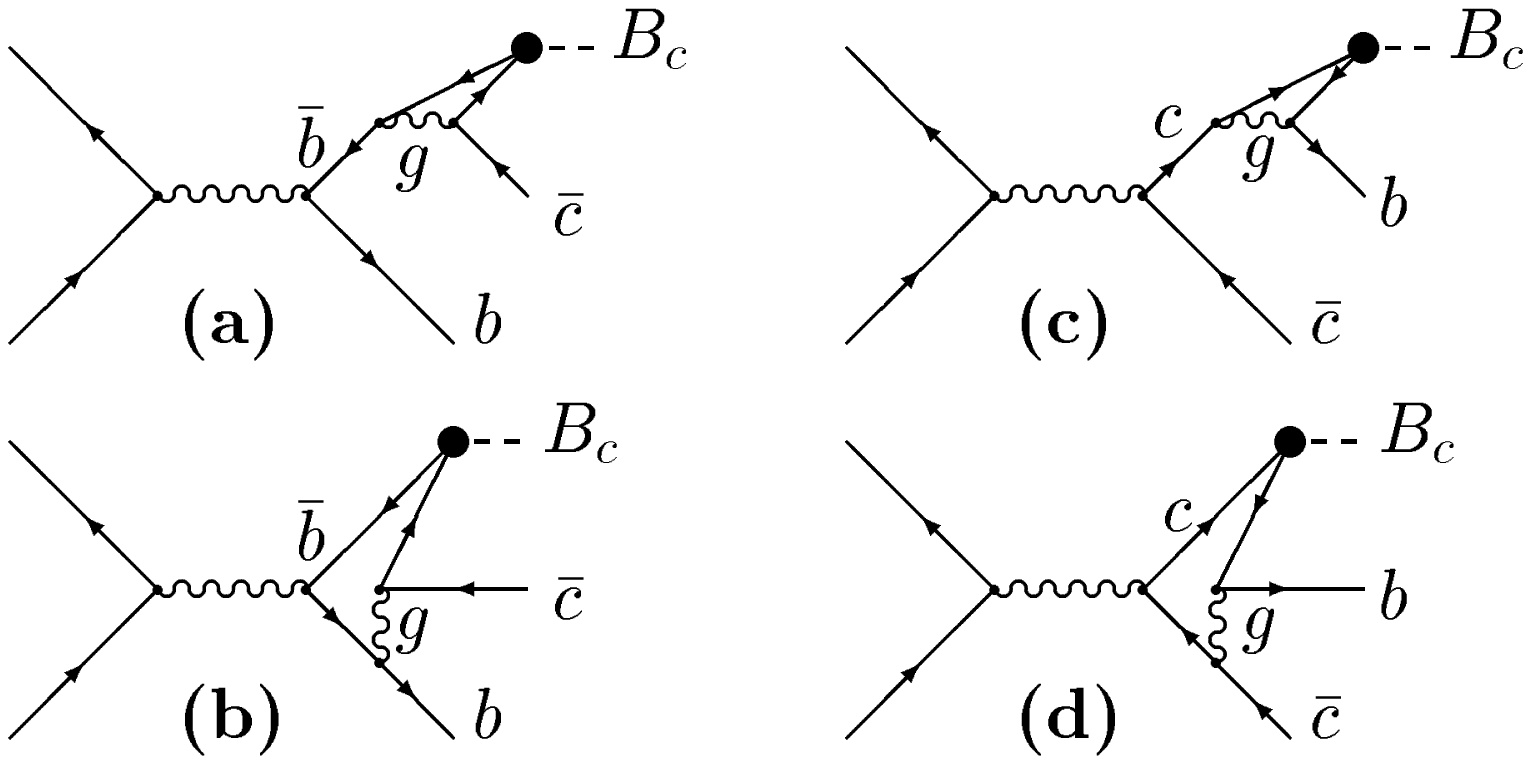}
}
\end{center}
\noindent
{\small\bf Fig.~1. }{\small\it
The lowest--order Feynman diagrams contributing to
$e^+e^- \rightarrow B_c b \bar{c}$.
}
}\end{minipage}
\hspace{0.5cm}
\begin{minipage}[t]{7.8cm} {
\begin{center}
\hspace{-1.7cm}
\mbox{
\epsfysize=7.0cm
\epsffile[0 0 500 500]{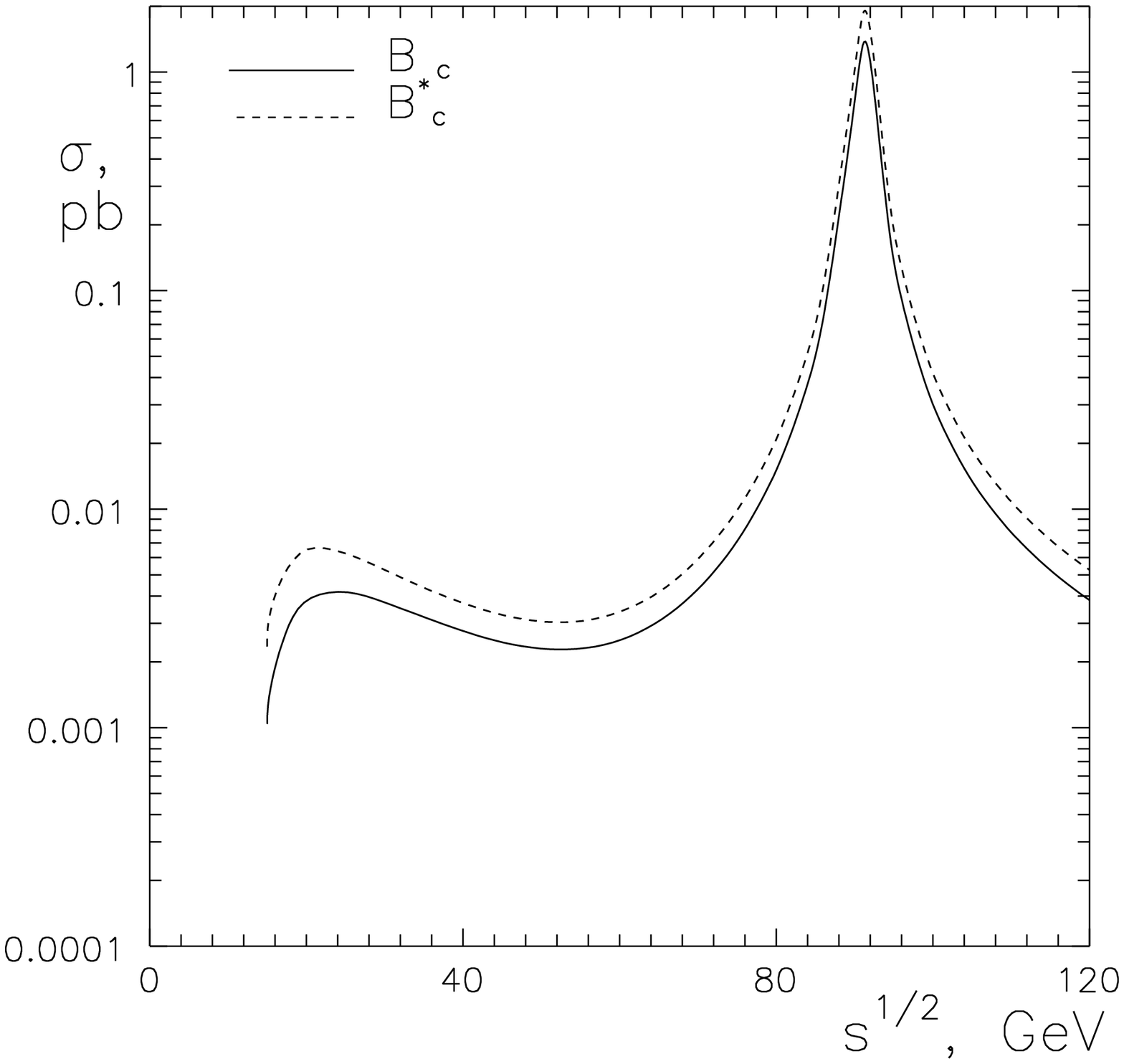}
}
\end{center}
\noindent
{\small\bf Fig.~2. }{\small\it
The total cross sections for $e^+e^- \rightarrow B_c^{(*)} b \bar{c}$
 versus the c.m. energy.
}
}\end{minipage}
\vspace*{0.5cm}

Following \cite{F1}, we write the differential $B_c$ cross
section as a convolution of the corresponding $b \bar b$ cross section and a
fragmentation function:
\bq
\label{bfrag}
d\sigma_{B_c}(E)=\int_0^1 dz d\sigma_{b\bar{b}}\left(\frac{E}{z},\mu\right)
D_{b\rightarrow B_c}(z,\mu),
\eq
where $\mu $ is the factorization scale and $z=2E_{B_c}/\sqrt{s}$.
For $\mu = O(m_{B_c})$ the
fragmentation function can be calculated directly in QCD perturbation theory
employing the nonrelativistic approximation sketched earlier. However, in
order to avoid large corrections proportional to $ln \frac{E}{\mu}$
in $d \sigma _{b\bar b}$, it is necessary to take $\mu = O(E)$ and to evolve
the fragmentation function $D_{\bar b \rightarrow B_c}$ from
$\mu \simeq m_{B_c}$
to $\mu \simeq E$ using Altarelli--Parisi equations. To leading
order, one has
\ba
\frac{\partial D_{\bar b \rightarrow B_c} (z, \mu )}
{\partial \, ln \, \mu ^2}
&=& \frac{\alpha _s(\mu)}{2 \, \pi} \, \int ^1_z \, \frac{dy}{y} \, P_{\bar b
\bar b} \left( \frac{z}{y} \right) D_{\bar b \rightarrow B_c} (y, \, \mu)
\nll
{\rm with}\ P_{\bar b \bar b} (x) &=& \frac{4}{3} \, \left[
\frac{1 + x^2}{(1 - x)_+}
+ \frac{3}{2} \, \delta (1 - x) \right] .
\ea
\ \vspace{1cm}\\
\begin{minipage}[tb]{7.8cm}{
\begin{center}
\hspace{-1.7cm}
\mbox{
\epsfysize=7.0cm
\epsffile[0 0 500 500]{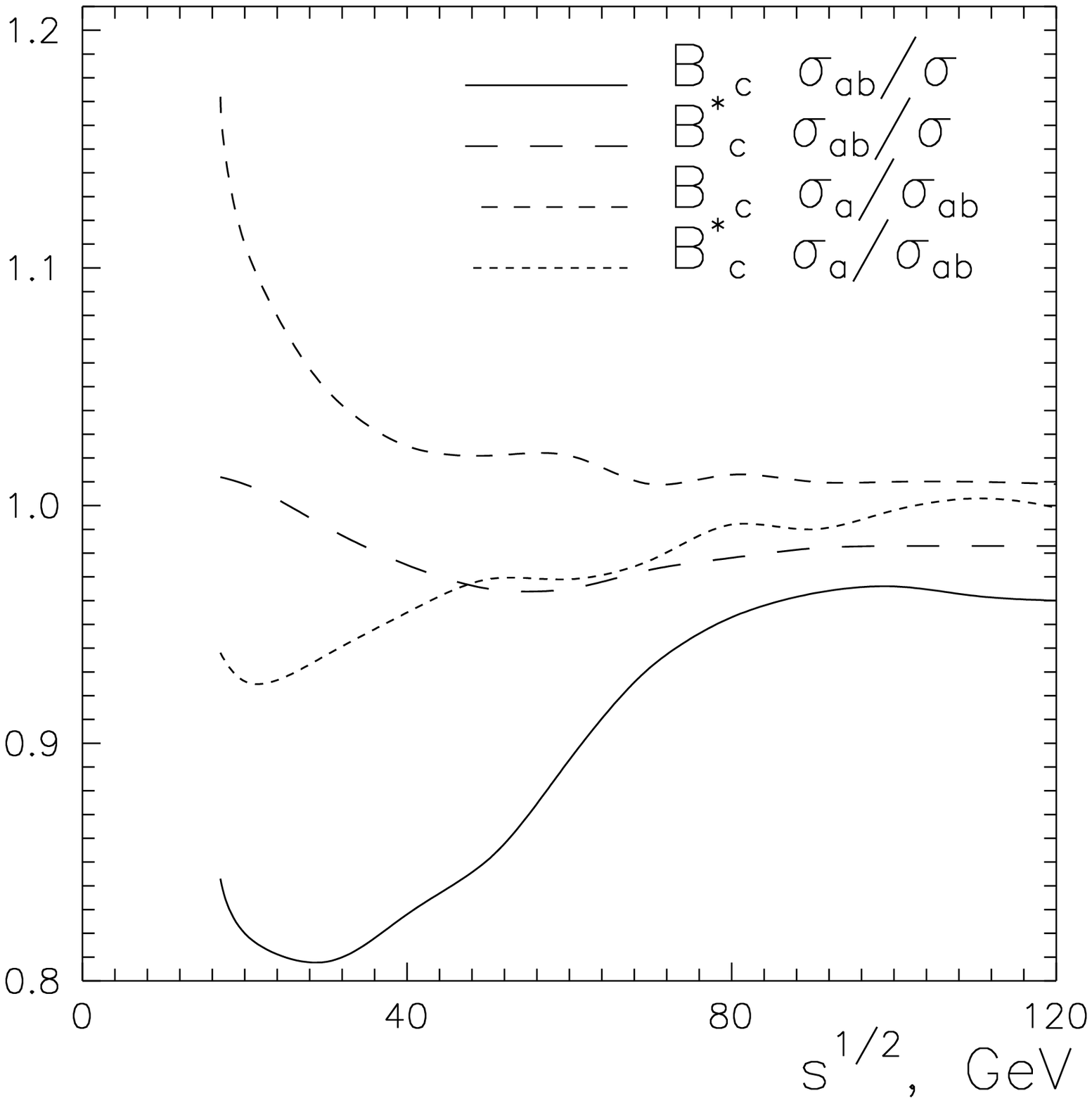}
}
\end{center}
\noindent
{\small\bf Fig.~3. }{\small\it
Cross section ratios calculated from Figs.~1a--d ($\sigma$), 1a and b
($\sigma_{ab}$), and 1a alone in planar gauge ($\sigma_a$).
}
}\end{minipage}
\hspace{0.5cm}
\begin{minipage}[tb]{7.8cm} {
\begin{center}
\hspace{-1.7cm}
\mbox{
\epsfysize=7.0cm
\epsffile[0 0 500 500]{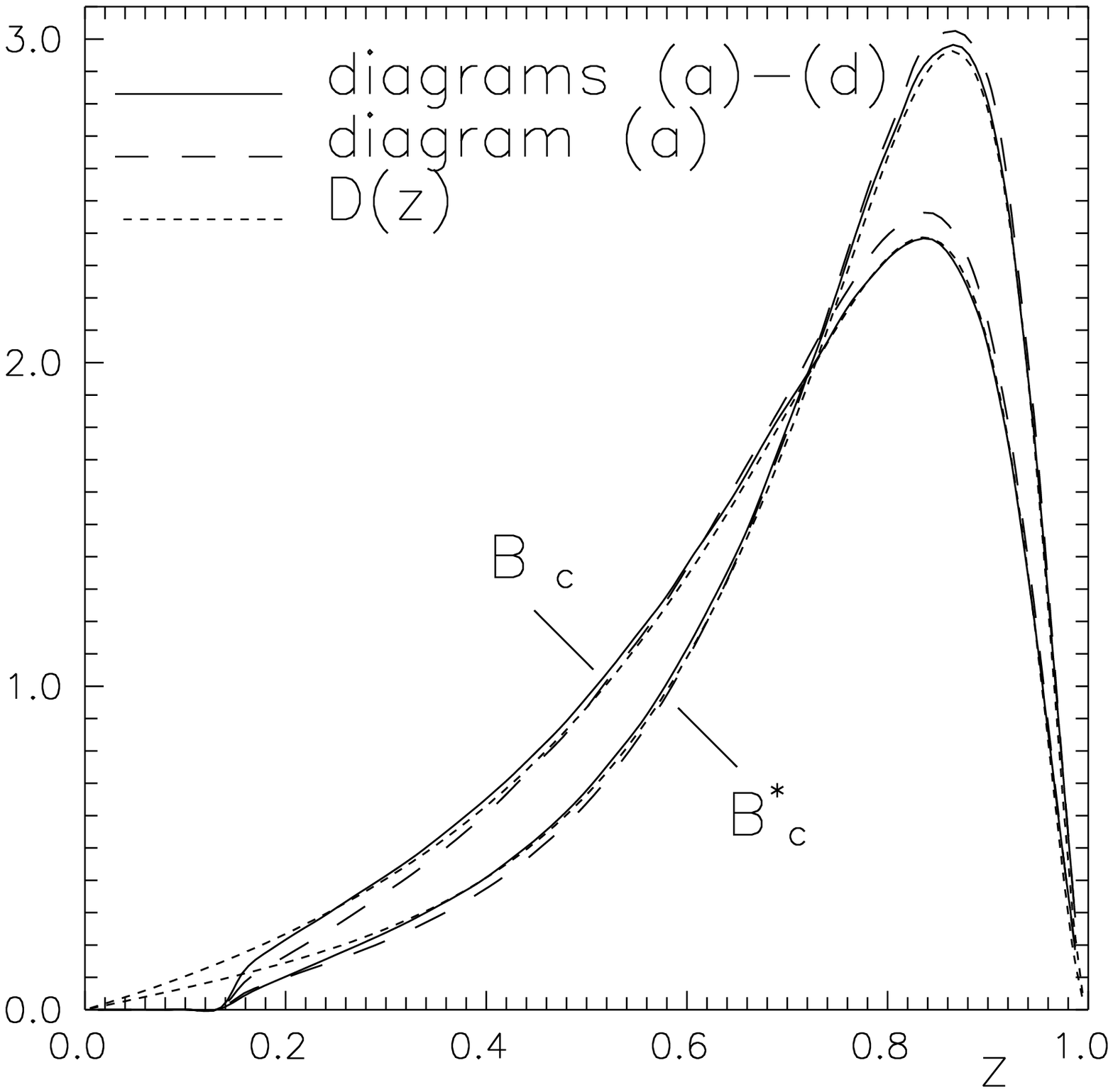}
}
\end{center}
\noindent
{\small\bf Fig.~4. }{\small\it
Energy distributions $\frac{\ d\sigma}{\sigma dz}$ calculated
from diagrams 1a--d (solid line), and from diagram
1a alone in planar gauge (dashed line) in comparison to the
fragmentation functions $D(z)$ given in (8), normalized to unity.
}
}\end{minipage}
\vspace*{0.5cm}

Fig.~4 shows the fragmentation functions at $\mu = m_{B_c}$
for $\bar b \rightarrow B_c$ and $\bar b \rightarrow B_c^*$ normalized to
unity in comparison to the distributions
$\frac{1}{\sigma} \frac{d \sigma}{dz}$ as calculated from
the complete set of diagrams in
Fig.~1, and from Fig.~1a only. With the exception of the threshold
behaviour at $z \rightarrow 0$, the direct calculation of
$D_{\bar b \rightarrow B_c^{(*)}}$
provides an excellent approximation. Evolution to
higher scales softens these distributions. This is exemplified in \cite{F1} for
$\mu = 10 m_{B_c}$. However, since
$\int^1_0 \, dz \, P_{\bar b \bar b}(z) = 0$,
the integrated fragmentation functions are not affected by
evolution. Consequently, the branching fractions for
$\bar b \rightarrow B_c^{(*)}  \, \bar c$ are universal and scale independent
numbers \cite{F1}:
\bq
BR\left(\bar{b} \rightarrow B_c^{(*)} \bar{c} \right) =
\left\{
\begin{array}{ll}
2.2 \cdot 10^{-4} & (B_c)\\
3.1 \cdot 10^{-4} & (B_c^*),
\end{array}
\right.
\eq
where $\alpha_s (m^2_{B_c}) = 0.2$ has been used. The fragmentation functions
for $P$--waves have been worked out in \cite{F2}.

It is instructive to compare the perturbative result:
\ba
\label{pertur}
D_{\bar{b}\rightarrow B_c^{(*)}}(z,\mu)
&=& N_{B_c^{(*)}}(\mu,m_c)\frac{z(1-z)^2}{(1-r_bz)^6} F_{B_c^{(*)}}(z,d),\nll
F_{B_c}(z,d)
&=& \frac{z^4}{16}(1+2d+2d^2+2d^3+d^4)-\frac{z^3}{12}(2+3d+10d^2+9d^3)\nll
& & +\frac{z^2}{6}(1+3d+17d^2)-3zd+1,\nll
F_{B_c^*}(z,d)
&=& \frac{z^4}{16}(5+10d+6d^2+2d^3+d^4)-\frac{z^3}{4}(8+7d+d^3)\nll
& & +\frac{z^2}{2}(9-3d+3d^2)-z(2+d)+1,
\ea
where $d=(m_b-m_c)/(m_b+m_c)$, and $r_b=m_b/(m_b+m_c)$,
with the phenomenological parametrization introduced in
\cite{P} for $D$ and $B$ meson fragmentation functions:
\bq
\label{peterson}
D(z)=N\frac{z(1-z)^2}{[(1-z)^2+\epsilon z]^2}.
\eq
Here, the shape parameter $\epsilon$ is considered to be of order
$m_q^2 / m_Q^2$,
$q$ and $Q$ denoting the light and heavy constituent quarks, respectively.
Interestingly, for $\epsilon \simeq m_c^2/4m_b^2$
the parametrization (\ref{peterson}) also provides a quite
reasonable description of the distributions (\ref{pertur}).
In addition to the effective shape
parameter
$\epsilon$, perturbation theory predicts the absolute normalization,
\bq
\label{norm}
N_{B_c^{(*)}}(\mu,m_c)=\frac{4 \alpha_s^2(\mu^2) f_{B_c^{(*)}}^2}{81m_c^2},
\eq
and the dependence on spin and orbital angular momentum
\cite{bcfrag,F1,falk,F2}.
%
%
\section{EVENT RATES AT THE Z POLE}
In the fragmentation picture it is easy to estimate the total number of
$B_c$ mesons produced at the $Z$ pole. One simply has to multiply the
branching fraction $BR \, (Z \rightarrow b \bar b) = 0.152$ with the
fragmentation probability of $b$ quarks
into   $B_c $ mesons. Since the
 excited states below the BD-threshold eventually decay into $B_c$ mesons,
they have also to be counted.
Adding up the $S$--waves, one obtains \cite{F1}
\bq
\label{bran}
BR(\bar{b}\rightarrow B_c\bar{c}X)\approx 9\cdot 10^{-4}.
\eq
One should, however, bear in mind the considerable uncertainties in this
number. These are mainly due to the parameters appearing in the normalization
factor (\ref{norm}), i.\ e.\ the scale $\mu $ in
$\alpha _s$, the quark
masses and the bound state wave functions or, equivalently, the decay
constants. Combined, there is certainly a factor two uncertainty in
(\ref{bran}) as can be seen from the varying predictions
in the
literature \cite{bcfrag,F1,falk}. From the branching fractions given above one
expects about
\bq
2700\ B_c/\bar{B}_c\ \ \mbox{per}\ \ 10^7\ Z.
\eq

About 20\% of these decay into
final states containing a $J/\psi $. In turn, 12\% of the $J/\psi $ resonances
decay
leptonically into $e^+ e^- $ or $\mu^+ \mu^-$. Hence, inclusively, one can
expect roughly 60 events per $10^7\ Z$ in the channel
$B_c\rightarrow J/\psi X;\ J/\psi\rightarrow l^+l^-.$
It should not be too difficult to observe such final states.
Unfortunately, the $B_c$ signature
will be obscured by a large background of $J/\psi $ events from other
sources. Most importantly, one expects more than
3000 events per $10^7\ Z$ from $Z \rightarrow b \bar b ;\ b/\bar b \rightarrow
J/\psi X;\ J/\psi \rightarrow l^+ l^-$,
where we have assumed an inclusive branching ratio of 1\% for $b \rightarrow
J/\psi $ X and the same for $\bar b $. This clearly shows that in order to
discriminate the $B_c $ events from the background one has to tag the charm
quark jet accompanying the $[\bar{b}c]$ bound states in
$Z \rightarrow b \bar b;
\ \bar b \rightarrow \nobody^{2S+1}L_J \bar c;\ ^{2S+1}L_J \rightarrow B_c X;
\ B_c \rightarrow J/\psi X$.

Alternatively, one may search for the $B_c$ meson in the particularly clean
detection channels
$B_c \rightarrow J/\psi l^+\nu _l$  and $B_c \rightarrow J/\psi \pi^+$.
Using the range of branching fractions given in
(\ref{branbc}), one can expect
$6 \div 16$ events for $10^7\ Z$ in the channel $l^+l^-l^+ \nu _l$
and a few events in the channels $l^+ l^- \pi^+$ and $l^+ l^- \rho^+$.
 These rates are small, but
maybe just sufficient for $B_c$ discovery.
%
\section{$B_c$ AT HADRON COLLIDERS}
Given the abundant production rates of $b$--quarks in hadronic collisions, the
prospects for detecting the $B_c $ mesons at the TEVATRON or LHC should look
much better. In addition to the small fragmentation
probabilities for $\bar b \rightarrow B_c \bar c$, these rates also
afford  small $B_c $ branching ratios.
The by far dominant subprocess is gluon--gluon fusion,
$gg \rightarrow B_c \bar c b $, with a small contribution coming also from
$q \bar q$ annihilation, $q \bar q \rightarrow B_c \bar c b$.

Unlike in the case of $e^+ e^-$ annihilation, only a minority of
diagrams possesses the
topology of fragmentation processes, e.\ g.\ $gg \rightarrow b \bar b $;
$\bar b \rightarrow B_c \bar c$. Correspondingly, heavy quark fragmentation
can only be expected to dominate $B_c$ production at large transverse
momenta. At smaller $p_T$ and for total cross sections, recombination
processes such as $gg \rightarrow b \bar b \, c \bar c$;
$\bar b c\rightarrow B_c $ will play an important role. The onset of the
asymptotic regime in $p_T$, where the fragmentation picture is valid, is
a priori not clear. We will return to this point in the next section.
%
\begin{center}\begin{tabular}{|c|r|c|r|}
\hline
  & $\sigma_{inc}$ & $\sigma_{dir}(${\small cuts}$)$ &
                                      $N_{inc}(${\small cuts}$)$\\
\hline
{\footnotesize TEVATRON}    &          &
                        \multicolumn{2}{c|}{\small $p_T>10\, GeV,\ |y|<1$}\\
$(25\,pb^{-1}) $         & 5.3\,nb & 0.13\,nb       & $1.6\cdot 10^4$\\
\hline
{\footnotesize LHC}         &          &
                        \multicolumn{2}{c|}{\small $p_T>20\, GeV,\ |y|<2.5$}\\
$(100fb^{-1})$         & 60\,nb  & 0.5\,nb               & $2.1\cdot 10^8$\\
\hline
\end{tabular}\end{center}
Table 2.
Cross sections and number of events at $\sqrt{s}=1.8\,TeV$ (TEVATRON) and
14(16)$\,TeV$ (LHC) for  inclusive ($inc$) and direct ($dir$) $B_c$ production.
\vspace{3mm}

In Table 2 we quote estimates for cross sections and production rates which
can be found in the literature. The total cross sections $\sigma _{inc}$ have
been calculated in \cite{PP1} taking into account the complete sets of
$O(\alpha_s^4)$--Feynman diagrams for the $gg$ and $q\bar q$ subprocesses,
and summing the individual cross sections for $1S$ and $2S$ waves. On the
other hand, the high-$p_T$ cross sections $\sigma _{dir}(cuts)$ have been
obtained in \cite{PP2} using an approximation analogous to (\ref{bfrag}),
and considering
only direct fragmentation into the $B_c$ ground state described by
(\ref{pertur}). Also applied are usual rapidity cuts. We
emphasize that the uncertainties in these predictions are even larger than
in the case of $e^+e^-$ annihilation. The main reason is the scale
ambiguity in the QCD coupling constant which enters here in fourth power.
This makes the evaluation of higher-order QCD
corrections very desirable.

As anticipated, even at high $p_T$ the production rates are large enough to
afford detection channels with branching ratios of order 1\% and below.
In fact, CDF has already recorded some candidate events \cite{MUE}.
A B-physics experiment
at LHC should certainly be able to investigate the $B_c$ system in some detail.
%
\section{PHOTON--PHOTON PRODUCTION -- A THEORETICAL STUDY CASE}
As pointed out, hadronic production of $B_c$ mesons involves fragmentation
and recombination mechanisms and is thus more involved than $B_c$
production by $e^+e^-$ annihilation. In order to learn about the relative
importance of fragmentation and recombination and their distinctive
features, we have investigated $B_c$ production in
$\gamma \gamma \rightarrow B_c b\bar c$. This process is simpler than
$gg$ fusion, but involves the same two mechanisms as can be seen from the
Feynman diagrams in Fig.~5. Moreover, it is interesting to find out whether
there is a chance to observe $B_c$ mesons in laser induced
$\gamma \gamma$ collisions
feasible at future linear collider facilities \cite{LC}.

In Fig.~6 we compare the cross section derived from the complete
set\footnote{Diagrams of type (I) with the photons
coupled to the $c$-quark line are not taken into account. They contribute an
extra 15\% to the total cross section \cite{klr}}
of $O(\alpha ^2 \alpha ^2_s)$ diagrams indicated in Fig.~5 with the result
obtained from the gauge invariant subset (I), which includes
the fragmentation diagrams.
Also shown is the cross section resulting from
a convolution of $\sigma (\gamma \gamma \rightarrow b \bar b)$
with the fragmentation function $D_{\bar b \rightarrow B_c}$ in analogy to
(\ref{bfrag}). We observe that the integrated cross section
is dominated by the recombination mechanism. This is not unexpected.
We also find agreement between the predictions of diagrams (I)
and simple $b-$quark fragmentation.
Apparently, there is a gauge in which the first diagram of set (I) dominates.
\ \vspace{1cm}\\
\begin{minipage}[t]{7.8cm}{
\begin{center}
\hspace{-1.7cm}
\mbox{
\epsfysize=7.0cm
\epsffile[0 0 500 500]{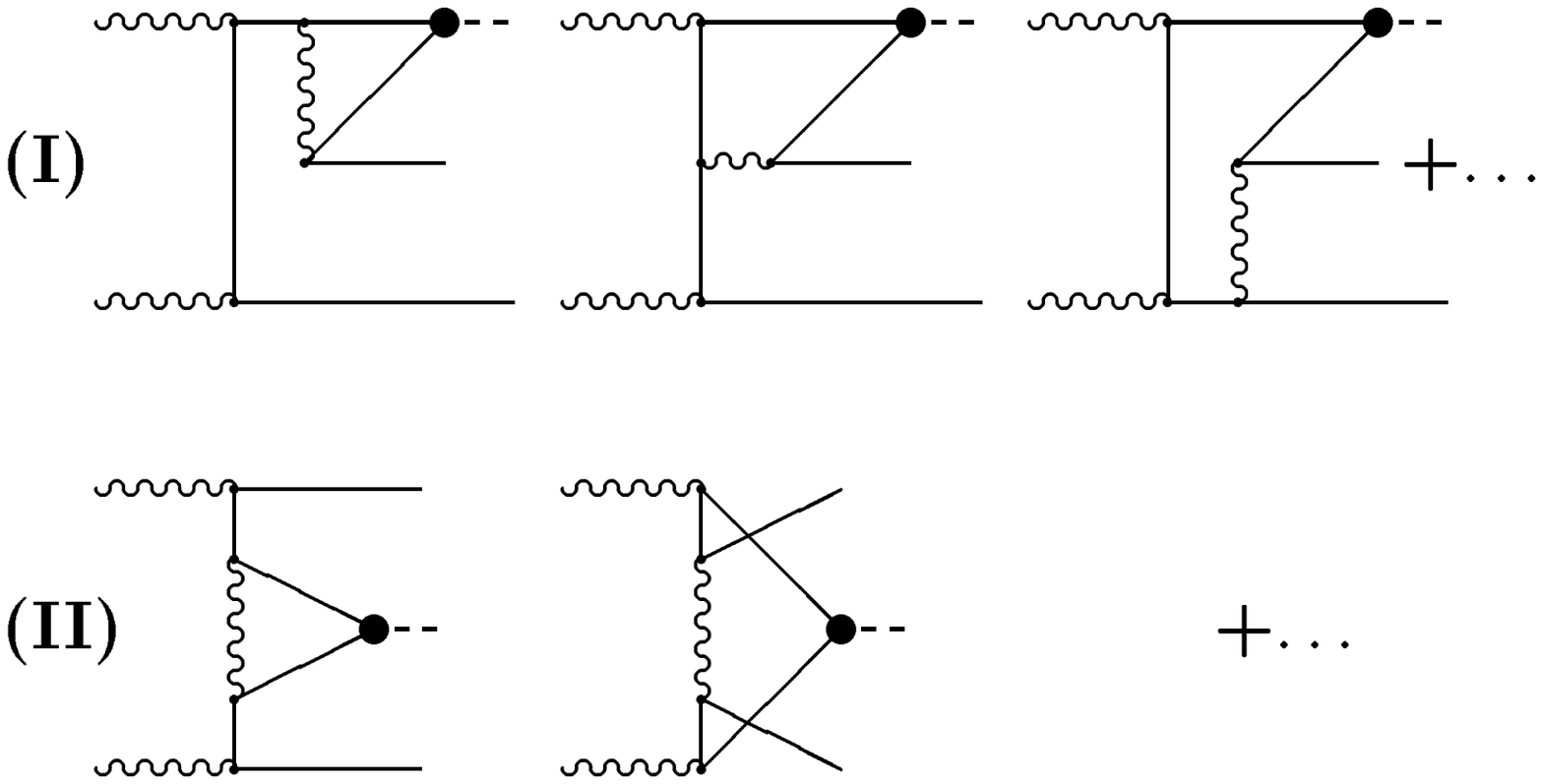}
}
\end{center}
\noindent
{\small\bf Fig.~5. }{\small\it
Different topologies of the lowest-order Feynman diagrams contributing
to $\gamma \gamma \rightarrow B_c c\bar{b}$.
}
}\end{minipage}
\hspace{0.5cm}
\begin{minipage}[t]{7.8cm} {
\begin{center}
\hspace{-1.7cm}
\mbox{
\epsfysize=7.0cm
\epsffile[0 0 500 500]{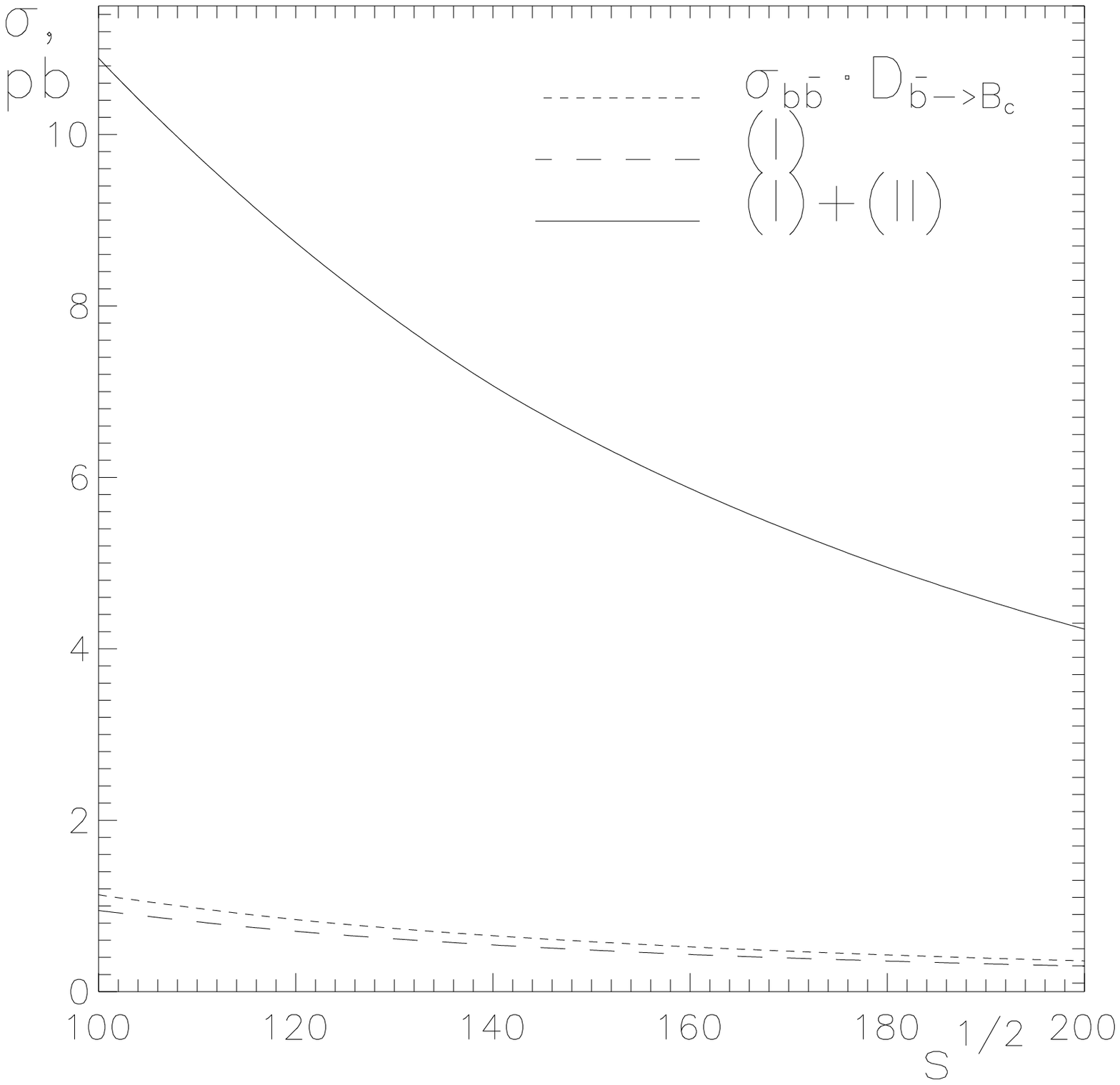}
}
\end{center}
\noindent
{\small\bf Fig.~6. }{\small\it
The total cross section for $\gamma\gamma \rightarrow B_c b \bar{c}$
 as function of the c.m. energy.
}
}\end{minipage}
\vspace*{0.5cm}

Furthermore, we have found that the energy distribution resulting from
subset (I) is very similar to the distribution predicted
by the pure fragmentation
description. This is illustrated in Fig.~7 together with the complete
energy distribution including also the  recombination processes (II).
The latter completely change the character of the energy spectrum giving
rise to rather soft $B_c$ production.

We have also compared the corresponding $p_{T}$ distributions. As
expected, the recombination processes mainly populate the low--$p_{T}$
region.
Nevertheless, their $p_T$ spectrum has a long tail towards high $p_T$.
For example, at $\sqrt{s}=100\,GeV$ and $p_T=20\,GeV$, the contribution
from the diagrams (I) and (II) are still equally important. Recombination dies
out completely only at $p_T\approx 40\,GeV$.
 These findings put some doubt
on the estimates in \cite{PP2} of $B_c$ production at high--$p_{T}$ in
hadronic collisions, where the fragmentation approximation is used at
$p_T\ll \sqrt{s}$.

In contrast to the energy and $p_T$ distributions, the angular distributions
resulting from fragmentation and recombination processes are quite similar
in shape.
All distributions are sharply peaked for $|\cos\theta |\rightarrow 1$,
with the
distribution derived from the complete set of diagrams in Fig.~5 being somewhat
flatter than the approximations from diagrams (I) and from
$d\sigma_{b\bar b}\otimes D_{\bar b\rightarrow B_c}$.

As a final remark, for $\sqrt{s} \geq 100 \, GeV$ we predict an
integrated cross section
below 12 fb. This shows that $\gamma \gamma$ colliders cannot be expected to
provide promising opportunities in $B_c$ physics.
%
\begin{center}
\begin{minipage}[t]{7.8cm} {
\begin{center}
\ \vspace*{0.5cm}\\
\hspace{-1.7cm}
\mbox{
\epsfysize=7.0cm
\epsffile[0 0 500 500]{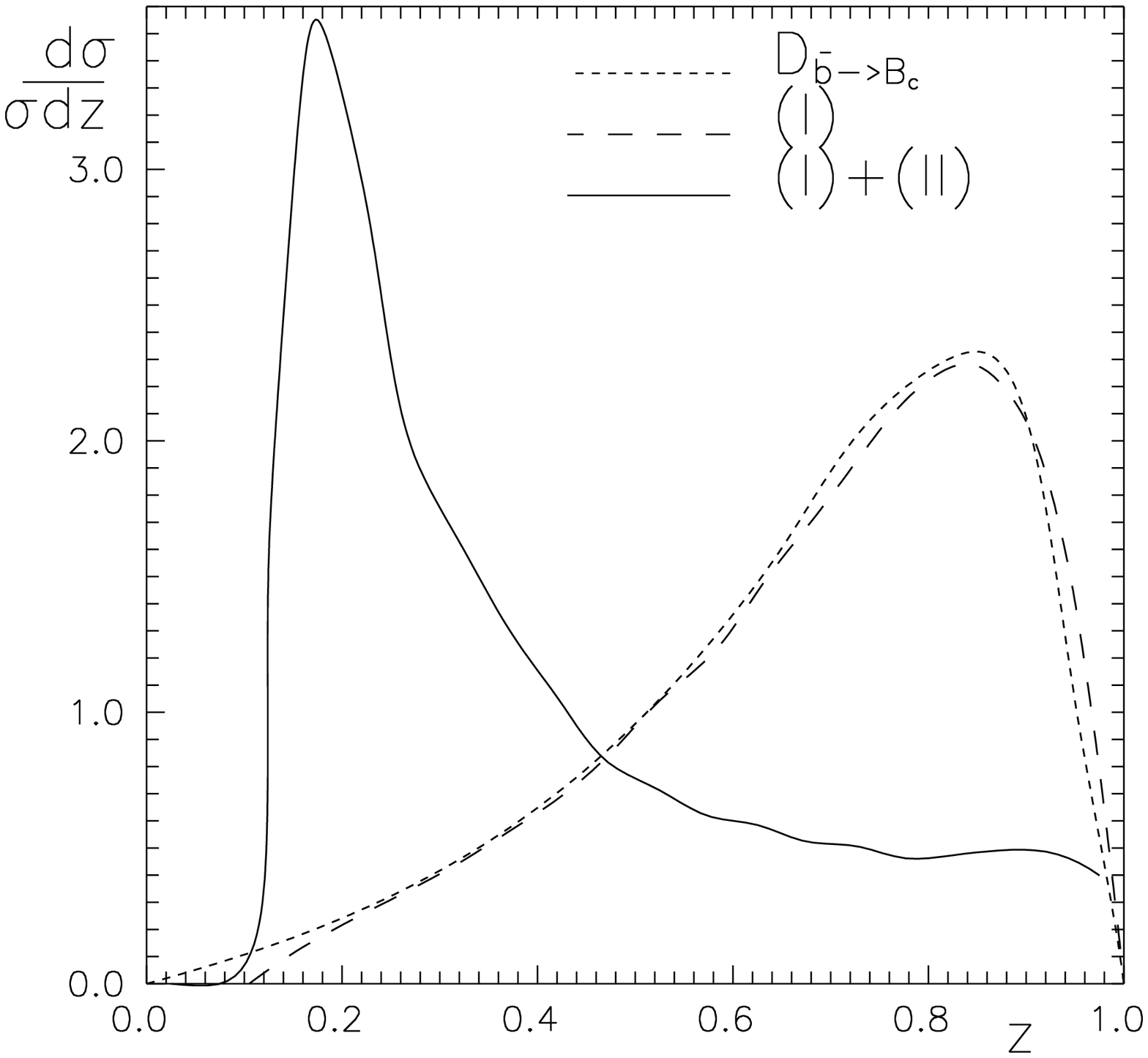}
}
\end{center}
}\end{minipage}
\end{center}
\noindent
{\small\bf Fig.~7. }{\small\it
Energy distributions
calculated from the diagrams (I) and (II) of Fig.~5, and from the subset (I)
alone in comparison
to the fragmentation function $D_{\bar b\rightarrow B_c}(z,\mu)$ given in (8),
normalized to unity.
}
\vspace{0.5cm}
%
\section{SUMMARY}
The $B_c$ system provides plenty of possibilities to investigate interesting
aspects of strong and weak interactions. Most importantly, the quarkonium
nature of these bound states allows to calculate and study also genuinely
nonperturbative
quantities such as total cross sections, decay amplitudes, transition form
factors and decay constants. Therefore, $B_c$ physics can serve as a testing
ground for quantitative theoretical approaches to confinement problems.
Practically, $B_c$ mesons are coming into experimental reach, may be at LEP,
likely at the TEVATRON, but most certainly at LHC.
%

\end{document}